\documentstyle[11pt,newpasp,twoside,epsf]{article}
\def\edcomment#1{\iffalse\marginpar{\raggedright\sl#1\/}\else\relax\fi}
\reversemarginpar

\begin{document}
\title{Galaxy Mergers: A Search for Chemical Signatures}
\author{Inese I.~Ivans}
\affil{Dept.~of Astronomy \& McDonald Observatory, Univ.~of Texas at Austin}
\author{Bruce Carney}
\affil{Dept.~of Physics \& Astronomy, Univ.~of North Carolina at Chapel Hill}
\author{Luisa de~Almeida}
\affil{Dept.~of Physics \& Astronomy, Univ.~of North Carolina at Chapel Hill}
\author{Chris Sneden}
\affil{Dept.~of Astronomy \& McDonald Observatory, Univ.~of Texas at Austin}

\begin{abstract}
We have gathered high resolution \'echelle spectra for more than two dozen 
high-velocity metal-poor field stars, including BD+80~245, a star previously 
known to have extremely low [$\alpha$/Fe] abundances, as well as G4-36, a new 
low [$\alpha$/Fe] star with unusually large [Ni/Fe].  In this kinematically 
selected sample, other chemically anomalous stars have also been uncovered.  
In addition to deriving the $\alpha$-element abundances, we have also analysed 
iron-group and $s$-process elements.  Not only does chemical substructure exist 
in the halo, but the chemical anomalies are not all the same within all 
elemental groups.
\end{abstract}

Recent kinematic and photometric evidence depicts an increasingly complicated 
picture of early galaxy formation, and of stellar debris left behind by an 
by an unknown number of passages and mergers by satellite systems of the 
Milky Way.  In addition to kinematic similarities, chemical ``signatures" may 
identify these merger/accretion events.  A number of very high-velocity 
metal-poor field stars have been discovered that possess very unusual 
abundance ratios of $\alpha$-elements (Mg, Si, Ca and maybe Ti) to iron.
The stars discovered to date all have large apogalacticon distances---these 
stars may have originated within a satellite galaxy or galaxies that 
experienced a different nucleosynthetic chemical evolution history than the 
Milky Way and which were later accreted by it.

Did the nucleosynthetic histories of the dwarf galaxies produce abundance
ratios different from those found in the halo of the Milky Way?  Preliminary results 
of chemical abundances of metal-poor stars in the Sagittarius dwarf spheroidal 
galaxy (Smecker-Hane et al.\ 1998) as well as in the Draco, Sextans and Ursa Minor 
dwarf galaxies (Shetrone et al.\ 2000), for the most part, seem to be similar to 
those of Galactic halo stars.  In the more metal-rich Sgr stars, however, 
Smecker-Hane et al.\ find subsolar elemental ratios indicative of an evolution 
consistent with enrichment from SNIa with declining SNII enrichment.  This is the 
kind of nucleosynthetic SNIa enrichment, as described by Unavane et al.\ 1996, one 
might expect from a system with a low star formation rate, along with possible 
recycling of nucleosynthetic products through successive stellar generations.  Dwarf 
galaxies today may not be representative of the dwarf galaxies of the past.  If 
mergers of dwarf galaxies at earlier times involved those less massive than the 
surviving satellites seen today, it is possible that the mergers left behind 
signatures of even less star formation and greater SNIa enhancements.

In figure~1, we illustrate the abundances of some of the low-$\alpha$ stars 
of abnormal abundances in the context of the rest of our sample and other 
stars of comparable metallicities from the literature.  For stars we have 
analysed in common with those in the literature, our abundances are in accord.

\begin{figure}
\plotone{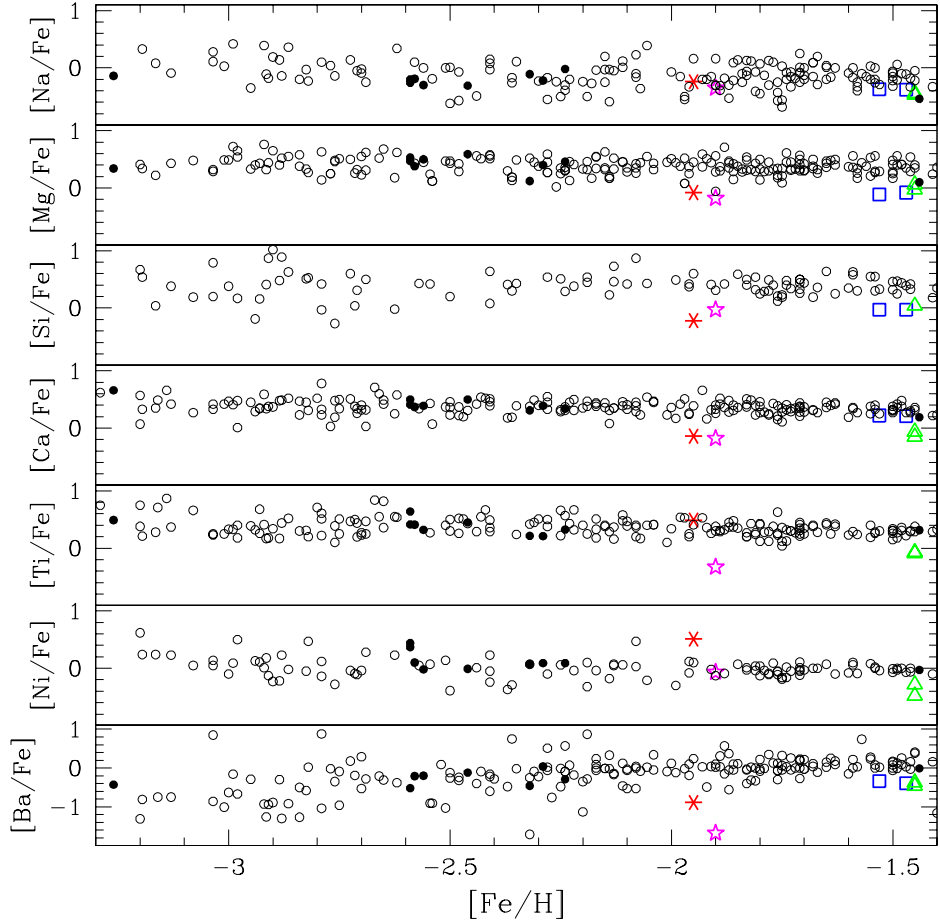}
\caption{The abundances of G4-36 ($*$) and BD+80~245 ($\star$) from this study,
HD139439/40 ($\Box$) from King 1997, and globular clusters Pal~12 and Rup~106 
($\triangle$) from Brown et al.\ 1997, along with some of our other stars 
($\bullet$) as well as results ($\circ$) from studies by Magain 1989, Gratton
\& Sneden 1991, McWilliam et al.\ 1995, 1998, Ryan et al.\ 1996, Carney et 
al.\ 1997, Nissen \& Schuster 1997, Hanson et al.\ 1998,  Stephens 1999,  
Burris et al.\ 2000, Fulbright 2000 and James 2000.  Plotted are abundances of
barium (an $s$-process element), nickel ($r$-process), titanium (sometimes 
$r$-process; sometimes an $\alpha$-element), calcium, silicon, magnesium (all 
$\alpha$-elements) and sodium.  These plots illustrate the unusual nature of 
some of the chemically anomalous halo stars (eg. most $\alpha$ and Ba 
abundances) {\it and} that these low-$\alpha$ stars do not exhibit the same 
anomalies in all element groups (eg. Ni and Ti).  Note the change in scale 
for the plot of [Ba/Fe] abundances.}
\end{figure}

We confirm that there appears to be a correlation between Na and Ni abundances 
with respect to iron in halo stars (Nissen \& Schuster 1997).  However, the trend of Na 
with Ni for the low-$\alpha$ stars seems to have a different slope and this is not 
understood.  We are currently exploring correlations between various abundance 
signatures with kinematics with the aim of unravelling the nucleosynthesis history that 
produced the unusual abundance ratios.

Chemical substructure in the halo as described here is not currently 
explained in models of galactic chemical evolution. With a 
sufficiently large representative halo sample, one can determine the fraction 
of chemically anomalous stars in the metal-poor halo and investigate their
origins.  By employing the kinematical information for these objects, we will
will also investigate whether the unusual [$\alpha$/Fe] ratios have been 
found in a large enough sample of stars to identify their origins.

\end{document}